\documentclass{iopart}

\usepackage[T1]{fontenc}

\usepackage{graphicx}
\usepackage{epstopdf}
\usepackage{amssymb}
\usepackage{tikz}
\usetikzlibrary{decorations.pathmorphing}
\usetikzlibrary{arrows,shapes}
\usepackage{subfigure}
\usepackage[urlcolor=blue,hyperindex,colorlinks,bookmarks=true,linkcolor=black,citecolor=black]{hyperref}
\usepackage[normalem]{ulem}
\usepackage[abs]{overpic}
\usepackage{color}
\usepackage{rotating}
\usepackage{enumerate}

\newcommand{\ket}[1]{\left|#1\right\rangle}
\newcommand{\bra}[1]{\left\langle#1\right|}
\newcommand{\braket}[2]{\left\langle#1\right|\left.#2\right\rangle}
\newcommand{\abs}[1]{\left|#1\right|}

\begin{document}

\title{Quantum Simulation of a Quantum Stochastic Walk}

\author{Luke C. G. Govia \footnote{Electronic address: govial@physics.mcgill.ca}\footnote{Current Address: Department of Physics, McGill University, Montreal, Quebec, Canada H3A 2T8}, Bruno G. Taketani, Peter K. Schuhmacher, and Frank K. Wilhelm}
\address{Theoretical Physics, Saarland University, Campus, 66123 Saarbr\"{u}cken, Germany}

\begin{abstract}
The study of quantum walks has been shown to have a wide range of applications in areas such as artificial intelligence, the study of biological processes, and quantum transport. The quantum stochastic walk, which allows for incoherent movement of the walker, and therefore, directionality, is a generalization on the fully coherent quantum walk. While a quantum stochastic walk can always be described in Lindblad formalism, this does not mean that it can be microscopically derived in the standard weak-coupling limit under the Born-Markov approximation. This restricts the class of quantum stochastic walks that can be experimentally realized in a simple manner. To circumvent this restriction, we introduce a technique to simulate open system evolution on a fully coherent quantum computer, using a quantum trajectories style approach. We apply this technique to a broad class of quantum stochastic walks, and show that they can be simulated with minimal experimental resources. Our work opens the path towards the experimental realization of quantum stochastic walks on large graphs with existing quantum technologies.
\end{abstract}

%\submitto{\NJP}
\maketitle

\section{Introduction}
\label{sec:intro}

Quantum walks offer a powerful paradigm for both studying, and harnessing quantum mechanics for computational applications \cite{Aharonov1993}. Quantum walks can be either continuous-time \cite{Farhi1998}, or discrete-time \cite{Aharonov2001,ambainis2001one}, and both types have been shown to be universal for quantum computation \cite{Childs2003,Lovett2010}. As a result, they can have computational advantages over classical algorithms, and quantum walks have been proposed for application in a variety of fields. Promising examples include machine learning \cite{Schuld2014c,Rebentrost2014}, artificial intelligence \cite{Cuevas,Briegel}, and photosynthetic excitation transfer \cite{Mohseni2008,Walschaers:2013aa}.

Extending on the idea of the continous-time quantum walk is the quantum stochastic walk (QSW), which combines the unitary evolution of a quantum walk with non-unitary stochastic evolution \cite{Whitfield2010}. This breaks time-reversal symmetry of the walker's evolution, and allows for the possibility of directed walks. Recently, it has been shown that when compared to their coherent counterparts, QSWs can have beneficial properties, such as speeding-up learning algorithms \cite{Schuld2014c,Cuevas}, or enhancing excitation transport \cite{Zimboras:2013aa,Mohseni2008}. Time-reversal symmetry can also be broken in chiral quantum walks \cite{Lu:2016aa}; however, these are completely coherent and therefore not QSWs.

One of the first discussed applications of quantum computing was the simulation of complex quantum systems \cite{Feynman:aa}, as a general purpose quantum computer can in principle simulate any quantum system \cite{Lloyd1073}. In recent times, purpose-built quantum simulators \cite{Buluta108} have found application in a range of fields, including quantum chemistry \cite{Lanyon:2010aa,Babbush:2014aa,OMalley2015}, relativistic quantum mechanics \cite{Gerritsma:2010aa,Lindkvist:2014aa}, and field theories \cite{Cirac:2010aa,Casanova:2011aa,Casanova:2011ab,Jordan1130}. Often, quantum walks can be thought of as simulation of quantum transport \cite{Mulken201137}.

In this work we introduce a technique for quantum simulation which uses a fully coherent quantum computer to simulate open system quantum evolution. Our simulation technique is the quantum analogue of the quantum trajectories technique used on classical computers \cite{Wiseman:2010fk}, and so we refer to it as quantum trajectories on a quantum computer (QTQC). We apply QTQC to the simulation of quantum stochastic walks; in essence performing a quantum simulation of what may be a quantum simulation itself.

We show that QTQC is computationally more efficient than a quantum trajectory simulation on a classical computer, due to the quantum nature of its implementation. In addition, we show in this work that QTQC simulations can be pieced together, allowing for simulation lengths that far exceed the coherence time of the physical simulator, and therefore, the maximum length of a fully coherent quantum simulation. As a result, QTQC simulation of interesting QSWs may be possible in the near future.

This paper is organized as follows. In section \ref{sec:QSW} we outline the formulation of a quantum stochastic walk, and briefly discuss why the required evolution is difficult to engineer physically \cite{Taketani2016}. In section \ref{sec:QSQW} we review the concept of quantum trajectories on a classical computer, and then describe the general scheme for quantum trajectories on a quantum computer. In section \ref{sec:QTQCSW} we apply QTQC to the simulation of a QSW. Finally, in section \ref{sec:conc} we make concluding remarks.

\section{Physical Realization of a Quantum Stochastic Walk}
\label{sec:QSW}

The quantum stochastic walks of Ref.~\cite{Whitfield2010} can be described by a Lindblad master equation, which generically takes the form
\begin{eqnarray}
\label{QSW}
\dot{\rho}= -i\left[\hat{H},\rho\right] + \sum_{k}\gamma_k{\left(\hat{L}_k\rho\hat{L}_k^{\dagger}-\frac{1}{2}\big\{\hat{L}_k^{\dagger}\hat{L}_k,\rho\big\}\right)},
\end{eqnarray}
where $\rho$ is the density operator of the walker, $\hat{L}_k$ are the Lindblad operators with $\gamma_k$ their associated incoherent transition rates, and $\hat{H}$ is the Hamiltonian describing the coherent part of the time evolution. The graph structure is encoded in the nonzero matrix elements of $\hat{H}$ (coherent edges) and the Lindblad operators with nonzero $\gamma_k$ (incoherent edges).

In this work we will be guided by proposals for QSWs solving various computational problems \cite{Cuevas,Briegel} in which the walker is restricted to the single excitation subspace of the graph. Therefore, each node can be described by a two-level system (qubit), with the excited state indicating the presence of the walker. For such a situation, the QSW can be described by the Lindblad master equation
\begin{eqnarray}
\dot{\rho}=-i\left[\hat{H},\rho\right] +\sum_{nm}\gamma_{nm}{\left(\sigma_m^{+}\sigma_n^{-}\rho\sigma_n^{+}\sigma_m^{-}-\frac{1}{2}\big\{\sigma_n^{+}\sigma_m^{-}\sigma_m^{+}\sigma_n^{-},\rho\big\}\right)},\label{QSWs}
\end{eqnarray}
where $\sigma_n^{+/-}$ is the raising/lowering operator for node $n$, and in general $\gamma_{nm}\neq\gamma_{mn}$. Crucially, the incoherent evolution of equation (\ref{QSWs}) also conserves the total excitation number in the graph, and as such the walker cannot be lost. It is important to point out that the QSWs considered here have only a positional degree of freedom, and are distinct from the open quantum walks of Refs.~\cite{Attal:2012aa,Sinayskiy:2013aa}, which contain both positional and internal (coin) degrees of freedom.

The QSW of equation (\ref{QSWs}) is a non-standard open system evolution, as incoherent excitation exchange occurs between the nodes, without local decay from the nodes into their environment, and possibly without local dephasing. As we have shown in Ref.~\cite{Taketani2016}, using standard two-body system-bath interactions and assuming an unstructured bath in the weak coupling limit, it is not possible to microscopically build a Lindblad equation of the form of equation (\ref{QSWs}). Heuristically, this can be understood to result from the fact that any incoherent evolution must arise from unitary coupling of the system to an environment, and that such coupling must take the form of local decay to the environment or local dephasing due to it. Therefore, one cannot avoid both of these local incoherent process and still have incoherent excitation exchange between nodes.

The restrictions found in Ref.~\cite{Taketani2016} can be circumvented with knowledge of the eigenspectrum of the graph Hamiltonian, and/or elaborate reservoir engineering. However, for QSWs of practical interest, the graph Hamiltonian will be sufficiently complicated that obtaining its eigenspectrum will be computational impractical on a classical computer. Also, while reservoir engineering can be useful in many circumstances \cite{Weimer2010,Murch2012,Fogarty2013,Govia:2015aa,Mourokh:2015aa,Metelmann:2015aa}, it requires an understanding and control of the environment that makes it impractical for implementing general, large-scale QSWs.

In this work we propose another way to circumvent the restrictions of Ref.~\cite{Taketani2016}, by simulating the desired QSW on a fully coherent quantum computer. In doing so, we use the QSW only as a quantum algorithm, and not a physical implementation. We will now introduce the concept of quantum trajectories on a quantum computer, which is a way of simulating general Lindblad open system evolution on a coherent quantum simulator.

\section{Quantum Simulation using Quantum Trajectories}
\label{sec:QSQW}

As we have just discussed, direct physical implementation of a system that evolves under the master equation of equation (\ref{QSWs}) poses a significant challenge. To circumvent this restriction, we propose simulation of equation (\ref{QSWs}) on a quantum computer using a quantum trajectories \cite{Wiseman:2010fk} style approach.

To begin, we consider the stochastic master equation unraveling of equation (\ref{QSW}), which is given by
\begin{eqnarray}
&\nonumber d\ket{\psi(t)} =\sum_k\Bigg[dN_k(t)\left(\frac{\hat{L}_k}{\sqrt{\left<\hat{L}^{\dagger}_k\hat{L}_k\right>(t)}}-1\right)\\
&+dt\left(\frac{\gamma_k\left<\hat{L}^{\dagger}_k\hat{L}_k\right>(t)}{2}-\frac{\gamma_k\hat{L}^{\dagger}_k\hat{L}_k}{2}-i\hat{H}\right)\Bigg]\ket{\psi(t)},
\end{eqnarray}
where $dN_k(t)$ is the stochastic increment for each Lindblad operator, for which the mean value is $E\left[dN_k(t)\right] = \bra{\psi(t)}\hat{L}^{\dagger}_k\hat{L}_k\ket{\psi(t)}$. We will first briefly review the quantum trajectories procedure used to simulate this equation on a {\it classical} computer and to find the density matrix of equation (\ref{QSW}). A full description of this technique can be found in Refs.~\cite{Jacobs:2006aa,Wiseman:2010fk}.

\subsection{Quantum Trajectories on a Classical Computer}
The approach to simulate quantum trajectories on a classical computer relies on a discretization of time and a separation between coherent and incoherent evolution. It consists of the following steps.

(1) {\it Coherent Evolution}: Starting at $t=0$, the system evolves under the unnormalized, non-Hermitian evolution
\begin{eqnarray}
\frac{d}{dt}\ket{\psi(t)} = -i(\hat{H}-i\hat{K})\ket{\psi(t)}, \label{eqn:Heff}
\end{eqnarray}
until a time $t_1$ such that $\braket{\psi(t_1)}{\psi(t_1)} = R_1$, where $R_1$ is a random number from the closed unit interval $[0,1]$. Here $K = \sum_k\gamma_k\hat{L}^{\dagger}_k\hat{L}_k/2$, and the random number $R_1$ is used to determine when an incoherent process (a quantum jump) occurs.

(2) {\it Incoherent Evolution or Quantum Jump}: The normalized expectation values
\begin{eqnarray}
E_k(t_1)= \frac{\gamma_k\bra{\psi(t_1)}\hat{L}^{\dagger}_k\hat{L}_k\ket{\psi(t_1)}}{\sum_k\gamma_k\bra{\psi(t_1)}\hat{L}^{\dagger}_k\hat{L}_k\ket{\psi(t_1)}}, \label{eqn:clexp}
\end{eqnarray}
are calculated, and used as weights to determine, via a second random number, which incoherent process $\hat{L}_k$ occurs at time $t_1$. Assuming that $\hat{L}_n$ is selected, then the state is updated via the rule
\begin{eqnarray}
\ket{\psi'(t_1)} = \frac{\hat{L}_n\ket{\psi(t_1)}}{ \bra{\psi(t_1)}\hat{L}^{\dagger}_n\hat{L}_n\ket{\psi(t_1)}},
\end{eqnarray}
which both applies the relevant jump operator, and renormalizes the state. The state $\ket{\psi'(t_1)}$ is then used as the new initial state.

Steps (1) and (2) are repeated, with new random numbers generated for each iteration, until the total simulation time $T$ is reached, producing an output state $\ket{\psi(T)}$ which corresponds to a single trajectory of the system evolution. An ensemble average of all possible trajectories gives the correct density matrix for a system evolving under the master equation of equation (\ref{QSW}), i.e.
\begin{eqnarray}
\rho(T) = {\rm E}\left[\ket{\psi(T)}\bra{\psi(T)}\right]. \label{eqn:DenMat}
\end{eqnarray}
Note that if $t_n > T$ no final incoherent jump is performed.

Quantum trajectory simulations on a classical computer of a system of Hilbert space dimension $D$ require $S$ trajectories to converge to an answer for the density matrix at time $T$. The runtime required is $O(SD^2)$, compared to the $O(D^4)$ runtime required for a numerical master equation solver \cite{Wiseman:2010fk}. For $S<D^2$ quantum trajectories on a classical computer is the more efficient technique \cite{Wiseman:2010fk}. It can also be beneficial if the density matrix is too large to store on a classical computer, but the state vector is not. Nevertheless, quantum trajectories is still an inefficient algorithm on a classical computer, as the Hilbert space dimension increases exponentially with the number of qubits.

\subsection{Quantum Trajectories on a Quantum Computer}
\label{sec:QTQC}
Practical implementations of the above protocol are inefficient on a classical computer, and hence would be limited to small graphs. To overcome this, one could envision implementation on a quantum computer. For this, the protocol requires the following modifications:

(1) {\it Coherent Evolution}: The evolution described by equation (\ref{eqn:Heff}) is in general nonphysical, and therefore is impossible to implement on a quantum computer with Hamiltonian evolution only. For time independent $\hat{H}$ and $\hat{K}$, the solution to equation (\ref{eqn:Heff}) is
\begin{eqnarray}
\ket{\psi(t)} = e^{-i(\hat{H}-i\hat{K})t}\ket{\psi(0)}.\label{eq:full_evolution}
\end{eqnarray}
If $\hat{H}$ and $\hat{K}$ commute, then this can be written as
\begin{eqnarray}
\ket{\psi(t)} = e^{-i\hat{H}t}e^{-\hat{K}t}\ket{\psi(0)},
\end{eqnarray}
and furthermore, if $\ket{\psi(0)}$ is an eigenvector of $\hat{K}$ with eigenvalue $\lambda$, then the solution for $\ket{\psi(t)}$ becomes
\begin{eqnarray}
\ket{\psi(t)} = e^{-\lambda t}e^{-i\hat{H}t}\ket{\psi(0)}. \label{eqn:HeffCom}
\end{eqnarray}
As can clearly be seen, equation (\ref{eqn:HeffCom}) is equivalent to the solution for evolution under the physical Hamiltonian $\hat{H}$ alone, up to a normalization factor $e^{-\lambda t}$. To render this nontrivial while obeying the condition that $K$ and $H$ commute, the eigenvalues of $K$ need to be degenerate. In section \ref{sec:QTQCSW} we will discuss the implications of this restriction in a specific example.

Therefore, we see that we can implement the coherent evolution step of the quantum trajectories algorithm on a quantum computer provided the following three conditions hold:
\begin{enumerate}
\item $\left[\hat{H},\hat{K}\right] = 0$,
\item The initial state, $\ket{\psi(0)}$, and the states at the start of each further coherent evolution step, $\ket{\psi'(t_i)}$, are all eigenstates of $\hat{K}$.
\item $\ket{\psi'(t_i)}$ is known at the end of each iteration, so that $e^{-\lambda_i t}$ can be calculated and used to determine the coherent evolution time $t_{i+1}$ for the next iteration, using $\braket{\psi(t_{i+1})}{\psi(t_{i+1})} = e^{-2\lambda_i t_{i+1}} = R_{i+1}$, with $R_{i+1}$ a random number from the unit interval. 
\end{enumerate}
Note that the second restriction can be lifted by simulating $e^{-\hat{K}t}\ket{\psi(0)}$ using a large ancilla system (see \ref{app:SNE}), and if this is the case, the formula to calculate the norm in restriction (iii) changes. Moreover, for large enough incoherent rates, under certain circumstances the first condition may be relaxed as the small average coherent time steps $t_i$ will justify a Suzuki-Trotter decomposition. See \ref{app:Commute} for further details.

(2) {\it Incoherent Evolution}: On a classical computer, the complete state $\ket{\psi(t)}$ after the previous coherent evolution is known, and so calculation of expectation values is simple. However, on a quantum computer the state is not known, and at each time $t_i$ either full state tomography must be performed to determine $\ket{\psi(t_i)}$, or each observable from equation (\ref{eqn:clexp}) must be measured sufficient times to obtain the relevant expectation values. Therefore, the first immediate problem with implementation of the incoherent step of the quantum trajectories algorithm on a quantum computer is efficient calculation of the expectation values $E_k(t)$ of equation (\ref{eqn:clexp}). 

As a result, each iteration step in a single trajectory must be run many times. The first $N-1$ times to generate sufficient measurement statistics so as to be able to determine the next incoherent quantum jump, and the $N$'th time to actually implement the next quantum jump, and continue on with the trajectory. This introduces considerable overhead to the protocol. If $S$ trajectories are required for convergence, a total of $O(N_{\rm jumps}N S)$ runs will be needed, where $N_{\rm jumps}$ is the average number of jumps per trajectory.

However, for certain classes of quantum jumps this overhead can be avoided. We will discuss one such case in the next section, where we consider a ``quantum trajectories on a quantum computer'' (QTQC) implementation of the class of quantum stochastic walks described by equation (\ref{QSWs}).

\section{QTQC of a Quantum Stochastic  Walk}
\label{sec:QTQCSW}

For the quantum stochastic walk of equation (\ref{QSWs}) we have
\begin{eqnarray}
\fl
\hat{K} &=\sum_{nm}\frac{\gamma_{nm}}{2} \sigma_n^{+}\sigma_m^{-}\sigma_m^{+}\sigma_n^{-} =\frac{1}{2}\sum_{n\neq m}\gamma_{nm}P^{(1)}_{n}\otimes P^{(0)}_{m}+\frac12 \sum_{n}\gamma_{nn}P_n^{(1)}
\end{eqnarray}
where $P^{(1)}_{n}$ is the projector onto the excited state of the qubit at node $n$ and $P^{(0)}_m$ the projector onto the ground state of the qubit at node $m$.

As we have a single walker on the graph, we can restrict our system to the single excitation subspace \cite{Pritchett:2010aa,Geller:2015aa}, and we use the notation $\ket{\phi_k}$ to indicate that the walker is in the $k$'th node of the graph. In the single excitation subspace the $\hat{K}$ matrix is diagonal and given by
\begin{eqnarray}
\hat{K}_{\rm SE} = \frac{1}{2}\sum_k \lambda_k \ket{\phi_k}\bra{\phi_k}, \label{eqn:KSE}
\end{eqnarray}
where $\lambda_k = \sum_n \gamma_{kn}$ is the total rate at which an excitation incoherently decays from node $k$ (into the other nodes). We consider a general Hamiltonian for a graph consisting of qubits coupled resonantly via the Jaynes-Cummings interaction, which in the single excitation subspace takes the form
\begin{eqnarray}
\hat{H}_{\rm SE} = \sum_{ij}g_{ij}\ket{\phi_i}\bra{\phi_j}, \label{eqn:HSE}
\end{eqnarray}
where the coupling strengths satisfy $\abs{g_{ij}} = \abs{g_{ji}}$ due to the symmetry of the Hamiltonian.

\subsection{Coherent Evolution}

\begin{figure}
\subfigure{\includegraphics[width = 0.45\columnwidth]{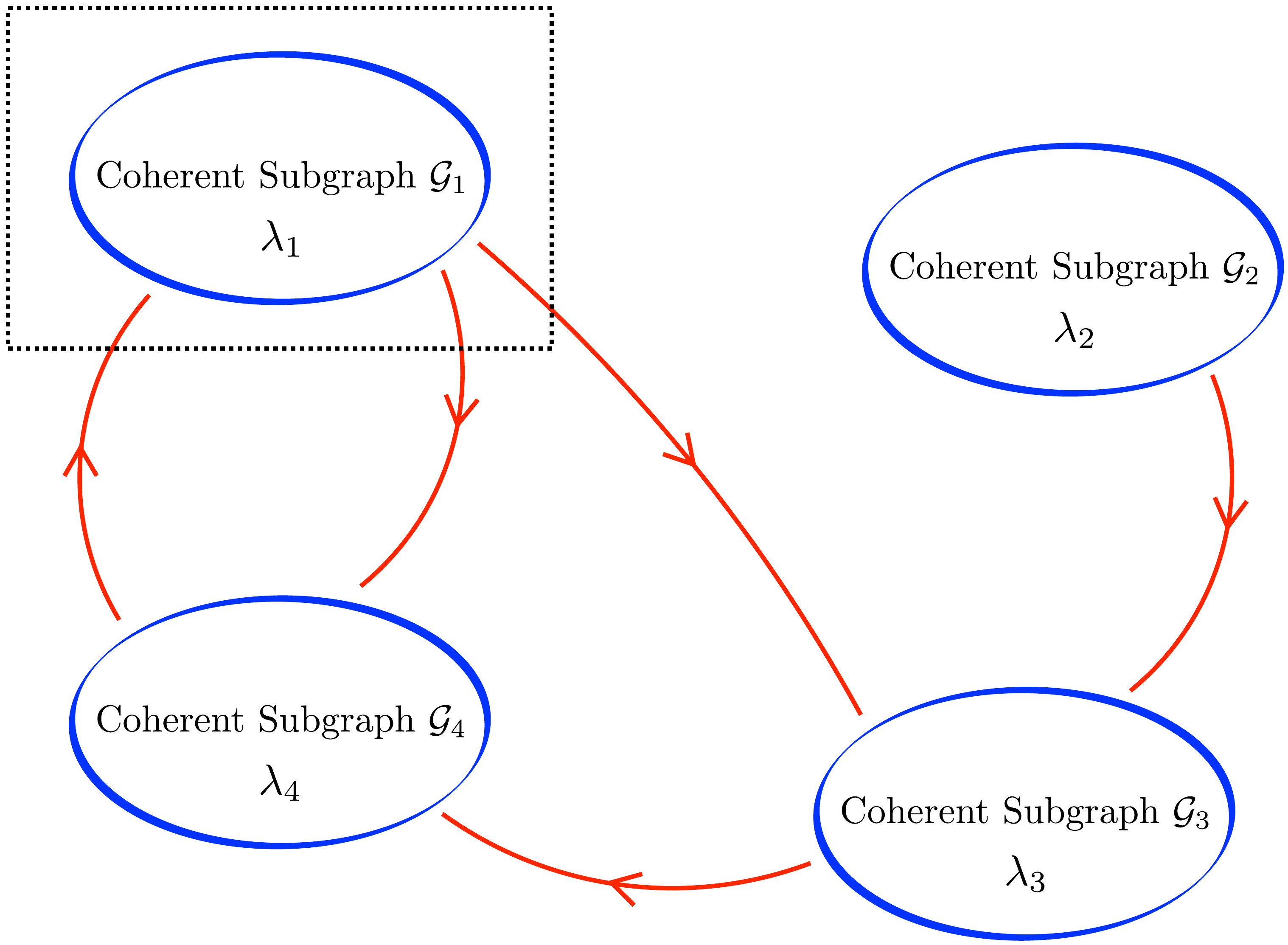}
\label{fig:Full}}
\subfigure{\includegraphics[width = 0.45\columnwidth]{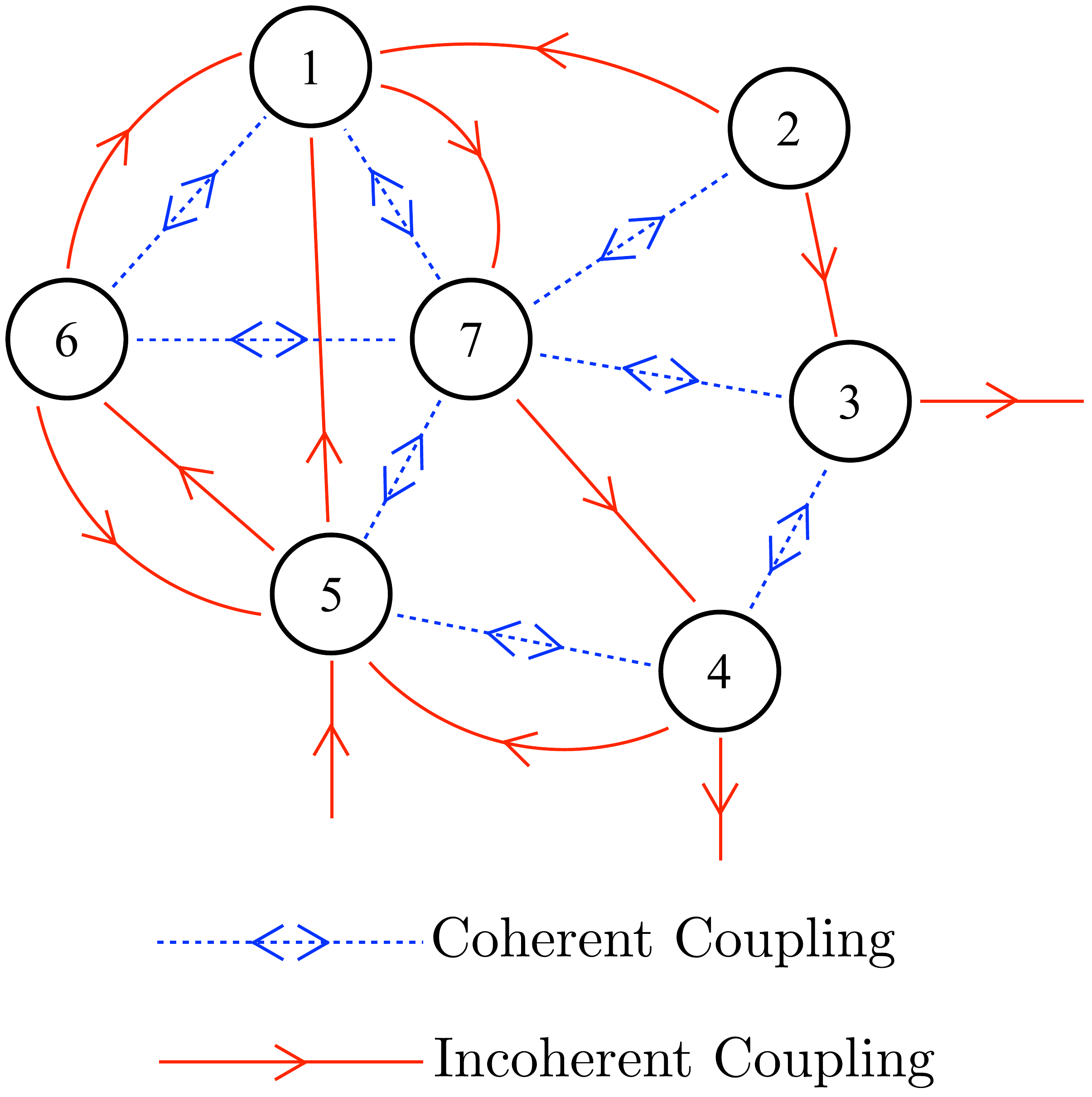}
\label{fig:Sub}}
\caption{\emph{Left Panel}: A sample graph, showing four coherently connected subgraphs linked by incoherent connections. \emph{Right Panel}: Zoom in of the dashed square in the left panel, an example of a single coherently connected subgraph with complicated network connectivity. The only restriction on the incoherent rates for this subgraph is that $\gamma_{17} + \gamma_{11} = \gamma_{21} + \gamma_{23} + \gamma_{22} = \gamma_{3*} + \gamma_{33}= \gamma_{4*} + \gamma_{45} + \gamma_{44}= \gamma_{56} + \gamma_{57} + \gamma_{55}= \gamma_{65} + \gamma_{61} + \gamma_{66}= \gamma_{74} + \gamma_{77}= \lambda_1$. Here $\gamma_{3*}$ and $\gamma_{4*}$ indicate the incoherent connections from nodes 3 and 4 that leave the subgraph. Notice that $\gamma_{*5}$, the incoherent connection entering the subgraph at node 5, plays no role in the definition of $\lambda_1$.}
\label{fig:Graph}
\end{figure}

To simulate this stochastic quantum walk with a QTQC approach, we still require that $\left[\hat{H}_{\rm SE},\hat{K}_{\rm SE}\right] = 0$. A simple calculation (see \ref{app:Commute}) shows that this is equivalent to the condition
\begin{eqnarray}
g_{nm}\left(\lambda_n-\lambda_m\right) = 0 \label{eqn:Con1}
\end{eqnarray}
for all nodes $n$ and $m$. What this means is that any two nodes with non-zero coherent coupling must have the same total incoherent decay rate $\lambda$. This is not in general true, and only a restricted set of graphs will satisfy this condition. An example of such a graph is shown in figure \ref{fig:Graph}.

Extending this condition to a coherently connected subgraph, one sees that all nodes of this subgraph must have the same total decay rate. The subgraph need not be completely connected, a single coherent connection between a new node and an existing subgraph is enough to enforce that the new node must have the same total decay rate as the nodes of the subgraph, see for example node 2 in the right panel of figure \ref{fig:Graph}. In addition, the only way to connect subgraphs with different $\lambda$'s is through a purely incoherent connection, as shown in the left panel of figure \ref{fig:Graph}.

In summary, the condition $\left[\hat{H}_{\rm SE},\hat{K}_{\rm SE}\right] = 0$ enforces that the total decay rate $\lambda$ must be the same for nodes that are coherently connected. We emphasize that this does not mean all $\gamma_{nm}$ must be the same {\em within} a coherently connected graph, only $\sum_k\gamma_{nk} = \lambda_n = \lambda$ must be constant for each node in the graph. Moreover, while all nodes have equal loss rates, the gain rates need not be equal, one can, for example, easily design a graph where an excitation can incoherently decay from, but never into, some given nodes. This implies that graphs with both source and sink nodes can be implemented. 

Furthermore, the coherent couplings between elements are unrestricted. As such, complicated connectivity networks are still possible, as demonstrated by the connectivity of the subgraph shown in the right panel of figure \ref{fig:Graph}. In addition, QSWs that satisfy the required criteria for QTQC simulation have already been shown to be advantageous in learning processes using neural networks \cite{Schuld2014c}.

In light of the previous discussion, we see that $\hat{K}_{\rm SE}$ takes the form
\begin{eqnarray}
\hat{K}_{\rm SE} = \frac{1}{2}\sum_i\sum_{k \in \mathcal{G}_i} \lambda_i \ket{\phi_k}\bra{\phi_k}, \label{eqn:Krest}
\end{eqnarray}
where $\mathcal{G}_i$ are the incoherently connected subgraphs of the graph $\mathcal{G}$, and $\lambda_i$ is the total decay rate of each node belonging to subgraph $\mathcal{G}_i$ (the left panel of figure \ref{fig:Graph} is an example of such a complete graph). To perform a QTQC simulation we also require that the initial state $\ket{\psi(0)}$ is an eigenstate of $\hat{K}_{\rm SE}$. From equation (\ref{eqn:Krest}) we see that this is satisfied provided $\ket{\psi(0)}$ is a superposition of nodes contained within a single subgraph $\mathcal{G}_i$ (assuming all $\lambda_i$ are distinct). In addition, we require that after each quantum jump the state $\ket{\psi'(t_i)}$ is an eigenstate of $\hat{K}_{\rm SE}$. Luckily, the form of the Lindblad operator, $\hat{L}_{nm} = \sigma_n^{-}\sigma_m^{+}$, ensures this is the case, as it localizes the excitation at node $m$ of the graph.

\subsection{Incoherent Evolution}

The fact that all Lindblad operators $\hat{L}_{nm}$ localize the walker to a single node is also beneficial for implementing the incoherent quantum jumps. This comes from the realization that the normalized expectation values
\begin{eqnarray}
\fl
\nonumber E_{nm}(t) = \frac{\gamma_{nm}\bra{\psi(t)} P^{(1)}_{n}\otimes P^{(0)}_{m}\ket{\psi(t)}}{\sum_{nm}\gamma_{nm}\bra{\psi(t)} P^{(1)}_{n}\otimes P^{(0)}_{m}\ket{\psi(t)}} =\frac{ \gamma_{nm} \abs{\braket{\psi(t)}{\phi_n}}^2}{2\bra{\psi(t)}\hat{K}_{\rm SE}\ket{\psi(t)}} \\ =\frac{ \gamma_{nm} \abs{\braket{\psi(t)}{\phi_n}}^2}{\lambda_n} =\frac{ \gamma_{nm} \abs{\braket{\psi(t)}{\phi_n}}^2}{\sum_k\gamma_{nk}} \label{eqn:probs}
\end{eqnarray} 
are equivalent to the probability that the excitation is in node $n$, given by $\abs{\braket{\psi(t)}{\phi_n}}^2$, multiplied by the probability the excitation decays into node $m$ from node $n$, given by $\gamma_{nm}/\sum_k\gamma_{nk}$.

If we measure the entire graph, we localize the excitation at a specific node $n$, which occurs with probability $\abs{\braket{\psi(t)}{\phi_n}}^2$. Next, using a random number and the weighted distribution $\gamma_{nm}/\sum_k\gamma_{nk}$ (which we know from designing the graph) we can determine into which node $m$ the excitation decays, and implement this transition. The net effect of this two-step process is that the probability of the transition from mode $n$ to mode $m$ is given by
\begin{eqnarray}
P_{nm} = \frac{ \gamma_{nm} \abs{\braket{\psi(t)}{\phi_n}}^2}{\sum_k\gamma_{nk}} = E_{nm}(t).
\end{eqnarray}

The choice of node $n$ is random due to the nature of quantum measurement, while the choice of node $m$ is random as we use a classical random number to choose $m$. Therefore, this hybrid quantum-classical probabilistic process samples randomly from the weighted distribution given by the expectation values $\{E_{nm}(t)\}$. In doing so, it correctly mimics the statistics of the QTQC simulation outlined in section \ref{sec:QTQC}, which allows the quantum jump to be implemented in a single shot with the correct statistics, without the large number of identical pre-runs normally required for a QTQC simulation.

It is important to point out that the simple form of the denominator of equation (\ref{eqn:probs}) is due to the fact that all coherently coupled nodes must have the same total decay rate $\lambda$. Therefore, since $\ket{\psi(0)}$ is an eigenstate of $\hat{K}_{\rm SE}$, then $\ket{\psi(t)}$ will also be an eigenstate of $\hat{K}_{\rm SE}$ as coherent evolution can only lead to a superposition of nodes which all have the same total decay rate.

\subsection{QTQC Protocol}

\begin{figure}
\begin{center}
\includegraphics[width = 0.5\columnwidth]{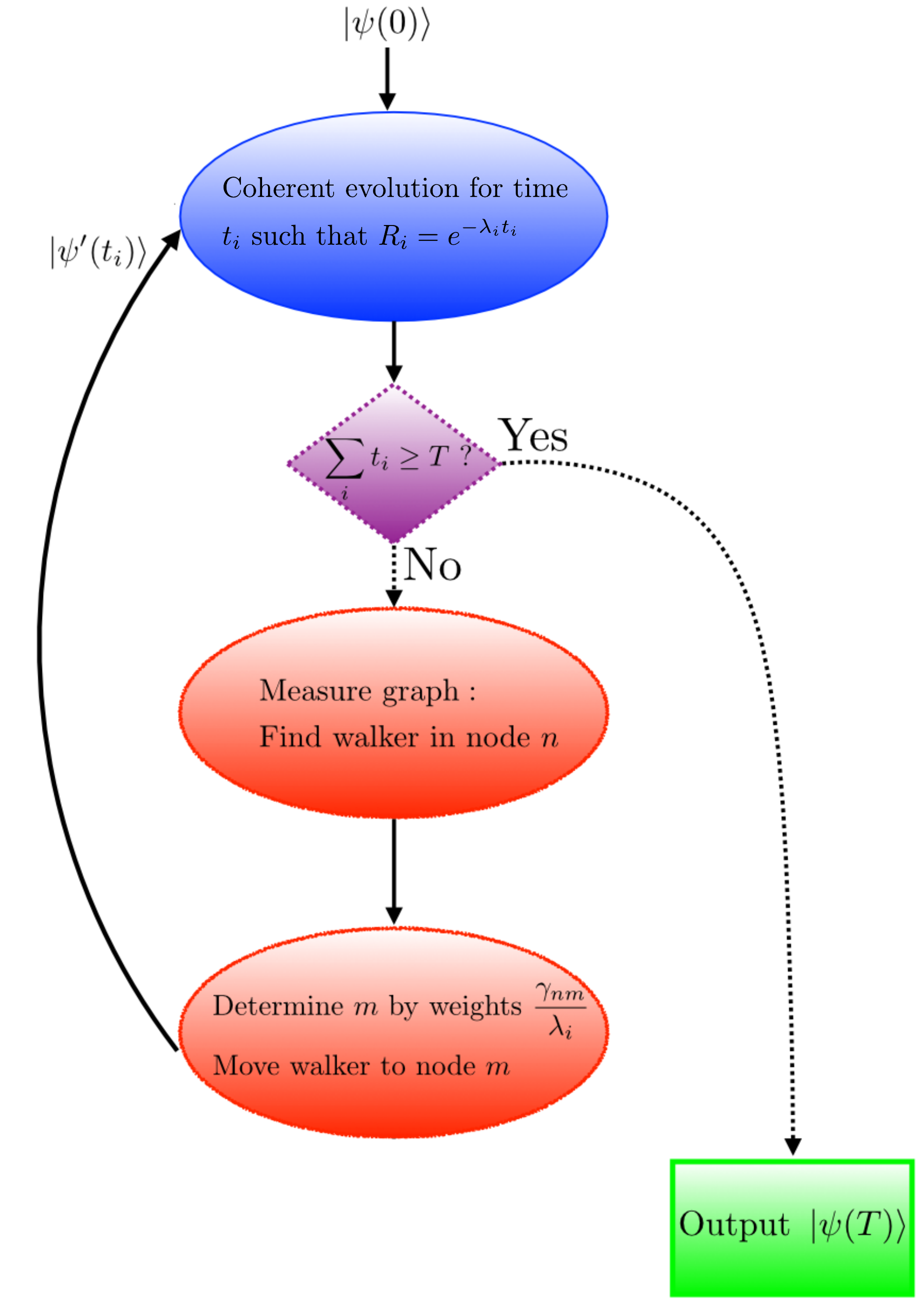}
\caption{Protocol to simulate a quantum stochastic walk via quantum trajectories on a quantum computer.}
\label{fig:protocol}
\end{center}
\end{figure}

Putting everything together, the full procedure for simulating a quantum stochastic walk using a QTQC protocol is as follows (also shown in figure \ref{fig:protocol}).
\begin{enumerate}[(1)]
\item The system starts in a state $\ket{\psi(0)}$ for which $\hat{K}_{\rm SE}\ket{\psi(0)} = \frac{\lambda}{2}\ket{\psi(0)}$. The system evolves coherently under $\hat{H}_{\rm SE}$ until a time $t_1$, such that $e^{-\lambda t_1} = R_1$, where $R_1$ is a random number from the unit interval.

\item The local population of the complete graph is measured (this measurement does {\it not} need to be quantum non-demolition). The walker is found in node $n$ with probability $\abs{\braket{\psi(t_1)}{\phi_n}}^2$, where $\ket{\phi_n}$ is the single excitation subspace state with the walker in node $n$.

\item A second random number is selected from the weighted distribution $\gamma_{nm}/\sum_k\gamma_{nk}$ to determine the destination node $m$, and the walker is re-initialized in node $m$. The graph is now in the state $\ket{\phi_m}$.

\item The above process is repeated, replacing $\ket{\psi(0)}$ with the localized state $\ket{\psi'(t_1)} = \ket{\phi_m}$, until the total evolution time $T$ is reached, with new random numbers being generated for each iteration. Due to the fact that the state at the beginning of each iteration is localized, it is guarantied that $\hat{K}_{\rm SE}\ket{\psi'(t_i)} = \frac{\lambda_i}{2}\ket{\psi'(t_i)}$ for each iteration. However, $\lambda_i$ may change between iterations, as the walker moves between subgraphs.
\end{enumerate}

Following this procedure, the quantum stochastic walk along any graph that satisfies equation (\ref{eqn:Con1}) can be simulated. The density matrix for the walker can be obtained using sufficient trajectories and state tomography.

\subsection{Resource Analysis and Scalability}

In this section we consider the resources required for a physical implementation of the QTQC simulation of a quantum stochastic walk. The number of trajectories required to accurately calculate the expectation value of an observable depends on the nature of the graph connectivity and on the observable, so general statements are difficult to make. However, the number of trajectories required for accurate results in a QTQC simulation will be the same as in a quantum trajectories simulation on a classical computer, as the statistical procedure is effectively the same.

As described before, a quantum trajectories simulation on a classical computer has a run time that scales as $O(SD^2)$, where $S$ is the number of trajectories required for converging results, and $D$ is the Hilbert space dimension of the system, while a numerical solution of the master equation has a runtime that scales as $O(D^4)$. Therefore, a trajectories simulation is more efficient for $S < D^2$ \cite{Wiseman:2010fk}. A QTQC simulation has the added benefit of requiring only $\log_2(D)$ qubits to simulate a quantum stochastic walk, that is, one for every node of the graph, unlike the classical simulation, which requires a number of bits that scales exponentially with the number of nodes.

The main resource requirements present in a QTQC simulation that do not have clear analogues in a quantum trajectories simulation on a classical computer are steps (2) and (3) of the protocol. These are full measurements of the graph to locate the walker, and moving the walker from one node to another, respectively. As a walker move only ever occurs after a graph measurement, we will only discuss the average number of measurements in a given trajectory. 

Sampling uniformly over the unit interval gives an expected $R_1$ of $\left<R_1\right> = 1/2$. Therefore, the average coherent evolution time before a quantum jump (and therefore a full graph measurement) satisfies
\begin{eqnarray}
e^{-\lambda t_{\rm avg}} = \left<R_1\right> = \frac{1}{2},
\end{eqnarray}
which implies that
\begin{eqnarray}
t_{\rm avg}  = \frac{\log\left(2\right)}{\lambda},
\end{eqnarray}
where $\lambda$ is the eigenvalue of $\hat{K}_{\rm SE}$ for the graph initial state. If $\lambda$ does not change after each measurement, then the average number of measurements per trajectory is simply given by
\begin{eqnarray}
\left<N_{\rm meas}\right> = \frac{T}{t_{\rm avg}},
\end{eqnarray}
where $T$ is the total time of the trajectory. When $\lambda$ does change this formula is no longer accurate, but by choosing the smallest $t_{\rm avg}$ (corresponding to the largest eigenvalue) one can calculate the ``worst case'' average number of measurements per trajectory.

However, it is not the number of required measurements that is most limiting to the size and complexity of the graph on which a QTQC simulation can be run. It is the required coherence time of the graph that is most limiting, as keeping large, strongly coupled networks of qubits coherent is a challenging experimental task. However, the coherence time is not given by the total length $T$ of a trajectory as one might na\"ively expect, but instead by the average time between quantum jumps $t_{\rm avg}$, as after each quantum jump the graph is ``reset'' into a definite (classically localized) state with no coherence. This potentially significant reduction in required coherence time increases the size of graphs on which a QTQC simulation could be run. In addition, the required coherence time actually decreases with the average number of measurements performed per trajectory, as the more measurements that are required, the stronger the incoherent process in the simulated quantum stochastic walk are in comparison to the coherent evolution.

It is also possible to use QTQC to simulate walks on graphs with more nodes than are experimentally feasible. Since the connections between coherent subgraphs are purely incoherent, they occur as quantum jumps, and the state of the walker is always confined to a single coherent subgraph. When a quantum jump between coherent subgraphs occurs, the experimental set-up needs only to be ``rewired'' so that it expresses the connectivity of the new coherent subgraph (and the excitation placed in the appropriate node). Therefore, the total number of physical nodes need only be as large as the largest coherently connected subgraph, provided the coherent coupling between physical nodes is tunable. One can image creating very large graphs built up of smaller coherent subgraphs in this way. Another approach one can envision would be to have different physical set-ups for each subgraph. As jumps effectively erase the memory of system, the statistical behaviour of each subgraph can be investigated independently and the final (complete) trajectories determined by connecting corresponding subgraph trajectories. This approach would require no rewiring of the set-up.

\section{Conclusion}
\label{sec:conc}

In this work we have introduced the concept of ``quantum trajectories on a quantum computer'' (QTQC), which is a quantum trajectories simulation of open system dynamics run on a quantum computer instead of a classical computer. As we have shown, QTQC cannot be used to simulate all Lindblad master equations, but when it can, it is more efficient than the classical simulation, owing to its quantum nature.

We have applied QTQC to simulating quantum stochastic walks (QSWs), a class of quantum algorithms that have many applications, including machine learning \cite{Schuld2014c,Cuevas}, and quantum transport \cite{Zimboras:2013aa,Mohseni2008}. We have found that using QTQC, one can simulate a restricted class of QSWs that still exhibit a flexible and rich graph topology. Examples of interesting QSWs that can be simulated with QTQC already exist \cite{Schuld2014c}. Additionally, for some graphs the restrictions can be lifted using ancillary systems and/or approximately lifted using a Suzuki-Trotter decomposition of the coherent evolution.

The coherence time of a QTQC simulator for a QSW need only be longer than the average time between quantum jumps, which can be many times shorter than the total simulation time. In addition, the QTQC simulator must only contain as many nodes as the largest coherently connected subgraph of the QSW, as QTQC trajectories can be pieced together. With these points in mind, QTQC simulation of a complex QSW on a large graph is likely achievable in the near future.

\label{sec:Conc}

\ack
The authors thank Frank Deppe for insightful discussions. Supported by the Army Research Office under contract W911NF-14-1-0080 and the European Union through ScaleQIT. LCGG acknowledges support from NSERC through an NSERC PGS-D.

\appendix

\section{Graph Restriction due to Commutation of $\hat{K}_{\rm SE}$ and $\hat{H}_{\rm SE}$}
\label{app:Commute}
To simulate a quantum stochastic walk using a QTQC protocol, it is required that the operators $\hat{K}_{\rm SE}$ and $\hat{H}_{\rm SE}$ commute. Using the form of these operators given in equations (\ref{eqn:KSE}) and (\ref{eqn:HSE}), we see that
\begin{eqnarray}
\fl
\left[\hat{H}_{\rm SE},\hat{K}_{\rm SE}\right] =  \frac{1}{2}\sum_{ijk}\Big(g_{ij}\lambda_k\ket{\phi_i}\braket{\phi_j}{\phi_k}\bra{\phi_k} - g_{ij}\lambda_k\ket{\phi_k}\braket{\phi_k}{\phi_i}\bra{\phi_j}\Big) \nonumber \\
=\frac{1}{2}\sum_{ijk}\Big(g_{ij}\lambda_k\delta_{jk}\ket{\phi_i}\bra{\phi_k} - g_{ij}\lambda_k\delta_{ki}\ket{\phi_k}\bra{\phi_j}\Big) \nonumber \\=\frac{1}{2}\sum_{ij}g_{ij}\left(\lambda_j-\lambda_i\right)\ket{\phi_i}\bra{\phi_j}, \label{eqn:Commute1}
\end{eqnarray}
where $\delta_{nm}$ is the usual Kronecker delta, and we have used the fact that the single excitation subspace states $\ket{\phi_n}$ are orthonormal. As the set of operators $\{\ket{\phi_i}\bra{\phi_j}\}_{ij}$ are mutually orthogonal, then each term in the sum in equation (\ref{eqn:Commute1}) must vanish independently, which leads to the graph restriction of equation (\ref{eqn:Con1}). 

The restriction $g_{ij}(\lambda_i-\lambda_j)=0$ does not need to precisely hold for the QTQC protocol to be applicable. If $(\lambda_i-\lambda_j)/\lambda_i\ll1$ for all $\{i,j\}$ for which $g_{ij}\neq0$, then the Suzuki-Trotter decomposition of equation (\ref{eq:full_evolution}) can be applied and the protocol can be used as an approximate solution.

\section{Simulating Nonphysical Evolution}
\label{app:SNE}

In this appendix we describe a protocol to simulate nonphysical evolution. Previous protocols have been developed to simulate specific nonphysical evolutions \cite{Casanova:2011aa}, and here we present a protocol to simulate the evolution $\ket{\psi(t)} = e^{-\hat{K}t}\ket{\psi(0)}$, where $K$ is a normal operator that is {\it not} skew-Hermitian, such that $e^{-\hat{K}t}$ is not a unitary matrix. 

However, as $\hat{K}$ is a normal matrix it is diagonalizable in its eigenbasis, which we shall label by $\left\{\ket{K_n}\right\}_{n=1}^D$, where $D$ is the dimension of the Hilbert space of the graph. We begin by expressing the initial state in terms of the eigenbasis of $\hat{K}$
\begin{eqnarray}
\ket{\psi(0)} = \sum_{n=1}^Dc_n\ket{K_n},
\end{eqnarray}
and it is clear that in this basis the final state is given by
\begin{eqnarray}
\ket{\psi(t)} = \sum_{n=1}^De^{-k_nt}c_n\ket{K_n},
\end{eqnarray}
where $k_n$ is the $n$'th eigenvalue of $\hat{K}$.

We introduce an ancillary quantum system of dimension $D$ spanned by the basis $\left\{\ket{\eta_n}\right\}_{n=1}^{D}$, which is initialized in a state $\ket{\Omega}$. Next, we perform the controlled entangling unitary
\begin{eqnarray}
\hat{U} = \sum_{n=1}^D \ket{K_n}\bra{K_n} \otimes \hat{U}_n,
\end{eqnarray}
where $\hat{U}_n\ket{\Omega} = \ket{\eta_n}$, such that the state of the joint system becomes
\begin{eqnarray}
\ket{\varphi} = \sum_{n=1}^D c_n \ket{K_n}\ket{\eta_n}.
\end{eqnarray}

Finally, we perform a measurement of the ancilla system in a basis that contains the state
\begin{eqnarray}
\ket{M} = \frac{1}{\sqrt{\mathcal{N}}}\sum_{n=1}^D e^{-k_n t} \ket{\eta_n},
\end{eqnarray}
where $\mathcal{N} = \sum_{n=1}^D \abs{c_n}^2e^{-2k_n t}$. When the outcome of the measurement is the state $\ket{M}$ the final state of the joint system is
\begin{eqnarray}
&\ket{\varphi'} = \frac{\mathbb{I}\otimes\ket{M}\braket{M}{\varphi}}{\sqrt{{\rm Tr}\left[\mathbb{I}\otimes\ket{M}\braket{M}{\varphi}\bra{\varphi}\right]}} \\ \nonumber&= \frac{1}{\sqrt{\mathcal{N}}}\sum_{n=1}^D c_n e^{-k_n t} \ket{K_n} \otimes \ket{M} = \frac{1}{\sqrt{\mathcal{N}}}\ket{\psi(t)}\ket{M}.
\end{eqnarray}
 As can be seen, the graph has been unentangled from the ancilla, and is now in the desired state $\ket{\psi(t)}$ (up to an irrelevant normalization factor).

The protocol is probabilisitic, as it succeeds only when the outcome of the ancilla measurement is $\ket{M}$, which happens with a probability given by the normalization factor $\mathcal{N}$. The longer the desired simulation time $t$, the smaller this factor, and therefore, the less likely the protocol is to succeed. 

In addition, this protocol creates the state $\ket{\psi(t)}$ for a single time $t$, and cannot simulate continuous time evolution under the nonphysical $\hat{K}$. For QTQC, one still needs to know the state $\ket{\psi'(t_i)}$ at the beginning of each coherent time step, in order to calculate the norm as a function of time, now given by the formula
\begin{eqnarray}
\braket{\psi(t_{i+1})}{\psi(t_{i+1})} = \bra{\psi'(t_{i})}e^{-2\hat{K}t}\ket{\psi'(t_i)} = \sum_{n=1}^D \abs{c_n}^2e^{-2k_n t}.
\end{eqnarray}

The protocol presented here is one example of a protocol to simulate $e^{-\hat{K}t}\ket{\psi(0)}$, and is neither meant to be optimal in any sense (resources, complexity, etc.), nor simple to implement in a physical system. It is only meant to highlight the fact that simulation of $e^{-\hat{K}t}\ket{\psi(0)}$ is possible in principle, and we anticipate that physical implementation of such a simulation will require extensive further theoretical and experimental work. 

One advantage of this approach based on quantum trajectories is that the dimension of the ancillary system scales with $D$, whereas general environmental representations require ancillary systems with dimension scaling with $D^2$ ~\cite{Bengtsson:2006}. We note that related protocols have recently been proposed in Ref.~\cite{Sweke2016}.

\section*{References}
\bibliographystyle{unsrt}
\bibliography{BibQuantumWalks}

\end{document}